\documentclass[12pt]{article}
\usepackage{amsfonts,amssymb,latexsym,amscd}
\title{\textbf{A tachyonic scalar field with mutually interacting components}}
\date{}
\author{Murli Manohar Verma\footnote{e-mail: sunilmmv@yahoo.com}\quad and\quad Shankar Dayal Pathak\footnote{e-mail: prince.pathak19@gmail.com} \\\emph{Department of Physics}, \emph{Lucknow University, Lucknow  226 007, India}
}

\begin{document}
\maketitle

\begin{abstract}

We investigate the tachyonic cosmological potential $V(\phi)$  in two different cases of the quasi-exponential expansion of universe and discuss various forms of interaction between the two components---matter and  the cosmological constant--- of the tachyonic scalar field,  which leads to the viable solutions of their  respective energy densities.  The distinction among the interaction forms is shown to appear in the $O_{m}(x)$ diagnostic.  Further,  the role of  the high- and low-redshift observations  of  the  Hubble parameter is discussed to determine the proportionality   constants and hence the correct form of matter--cosmological  constant   interaction.

\end{abstract}

matter; cosmological constant.


\pagebreak

\section{Introduction}

A bunch of cosmological observations over the past decade, such as SNe type Ia \cite{a1,b1}, Cosmic Microwave Background Radiation(CMBR) \cite{b2,d1}, Baryon Acoustic Oscillations(BAO) \cite{e1,f1} in galaxy surveys etc.  show that nearly 73\% of the total contents in the universe is composed of an exotic component which,  often termed as dark energy, is held responsible for the accelerated expansion of the universe at the present epoch.

A class of scalar fields is one of the promising candidates of dark energy \cite{i1,j1,k1}. Among itself, the  tachyonic scalar field arising from string theory \cite{n1} (for different reasons in our context)  has been  widely used in literature \cite{l1,m1,o1}. For us here, one of the main  reasons for this  predilection   is  that the Lagrangian adopted in  the tachyonic scalar field is relativistic which is logically more appealing than its non-relativistic counterpart usually adopted in quintessence scalar field.

The energy density of the tachyonic scalar field includes two components. If we take one component as the pressureless dark matter with $w_{m}=0$ then the other one is found to behave as the cosmological constant with $w_{\lambda}=-1$. We take the dark matter component as inclusive of the baryonic contributions having the same equation of state while any contributions arising from radiation would be almost negligible in a matter dominated universe. Several workers have investigated the cosmological behaviour of the the tachyonic scalar field \cite{l1,m1,o1}, and have found forms of potential for power law expansion and pure exponential expansion. Since the present observations have not yet been able to fix the exact form of the evolution of the  expansion factor\footnote{Even though the present  supernovae  observations \cite{a1,b1}  seem to indicate an accelerated expansion of the universe, yet the exact form of the expansion has still  not been  unambiguously  ascertained. It is possible that we have already entered another  inflationary phase.}, we attempt to study the alternatives based on superposition of these forms.

In section 2 of this paper,  we find the potential $V(\phi)$ of the  tachyonic scalar field for quasi-exponential expansion of the universe as a product and sum of power law and exponential expansion.  In section 3  we study the mutual interaction between components of the tachyonic scalar field (\emph{viz}., dark matter and the cosmological constant) to determine $\rho_{m}(a)$ and $\rho_{\lambda}(a)$.  Section 4  is devoted to discuss the effects of  the interaction in  $O_{m}(x)$  diagnostic. The calculation of proportionality constants in  various  forms of the interaction term $Q$, using the observed values of  Hubble parameter at different redshifts,  is presented in section 5. This is followed by our concluding remarks in section 6 with a further possible  evaluation of the interaction to constrain the structure formation.

\section{Potentials  $V(\phi)$  of the tachyonic scalar field}
The relativistic Lagrangian proposed \cite{n1} for the tachyonic scalar field $\phi$
\begin{eqnarray}L=-V(\phi)\sqrt{1-\partial_{i}\phi\partial^{i}\phi}\label{n1}\end{eqnarray}
and the stress energy tensor
\begin{eqnarray} T^{ik}=\frac{\partial{L}}{\partial{(\partial_{i}\phi)}}\partial^{k}\phi-g^{ik}L\label{n2}\end{eqnarray}
give the energy density and pressure as
\begin{eqnarray} \rho=\frac{V(\phi)}{\sqrt{1-\partial_{i}\phi\partial^{i}\phi}}; \qquad P=-V(\phi)\sqrt{1-\partial_{i}\phi\partial^{i}\phi}\label{n3}\end{eqnarray} respectively, which for spatially homogeneous field, reduce to
\begin{eqnarray}\rho=\frac{V(\phi)}{\sqrt{1-\dot{\phi^{2}}}} ;\qquad P=-V(\phi)\sqrt{1-\dot{\phi^{2}}}\label{n4}\end{eqnarray}
where an overdot denotes the derivative with respect to time (as in the rest of this paper). Here $\rho$ and $P$ in (\ref{n4}) can be decomposed into an arbitrary number of components, each with its own equation of state.  However, in the post-recombination era  (subsequent to $z\sim1000$), and more effectively towards the present epoch $(z\rightarrow 0)$, the universe is largely dominated by matter and dark energy (strongly represented by the cosmological constant). Therefore, we are motivated to consider only two components matter and the cosmological constant which fit in so well that if one component is matter with $w_{m}=0$, then the other one turns out to be the cosmological constant with $w_{\lambda}=-1$. Thus considering
\begin{eqnarray}\rho=\rho_{1}+\rho_{2} ;\qquad P=P_{1}+P_{2}\label{n5}\end{eqnarray}
we get the components from (\ref{n3}) as
\begin{eqnarray}\rho_{1}=\rho_{m}
=\frac{V(\phi)\partial_{i}\phi\partial^{i}\phi}{\sqrt{1-\partial_{i}\phi\partial^{i}\phi}} ;\qquad P_{1}=P_{m}= 0\label{n7}\end{eqnarray}

and \begin{eqnarray}\rho_{2}=\rho_{\lambda}=V(\phi)\sqrt{1-\partial_{i}\phi\partial^{i}\phi} ;\qquad P_{2}=P_{\lambda}=-V(\phi)\sqrt{1-\partial_{i}\phi\partial^{i}\phi}\label{n8}.\end{eqnarray}

We consider our universe in a quasi-exponential phase (product of power law and exponential, or linear combination of power law and exponential) with the scale factor respectively given as
\begin{eqnarray}a(t)=a_{0}t^{n}exp(\alpha t)\label{n10}\end{eqnarray}
\begin{eqnarray}a(t)=At^{n}+B exp(\alpha t)\label{n11}\end{eqnarray}
where $a_{0}, \alpha, A $ and $B$ are all constants.

The Friedmann equations give (for $k=0$ universe)
\begin{eqnarray}\left(\frac{\dot{a}}{a}\right)^{2}=\frac{\rho}{3M^{2}_{Pl}} ;\qquad   \left(\frac{\ddot{a}}{a}\right)=-\frac{(\rho+3P)}{6M^{2}_{Pl}}\label{n12}\end{eqnarray}
where $M_{Pl}=(8\pi G)^{-1/2}$ is the reduced Planck mass  with $\rho(\phi)$ and $P(\phi)$ given by (\ref{n4}).

The conservation of field energy gives

\begin{eqnarray}\dot{\rho}=-3H(1+w)\rho\label{n13}\end{eqnarray}
with the equation of state parameter $w=P/\rho$ and   $1+w=\dot{\phi}^{2}(t)$.  From (\ref{n12}), we also have $\dot{\rho}/\rho=2\dot{H}/H$.
 Using (\ref{n4}), (\ref{n12}) and (\ref{n13}) we can calculate $\phi(t)$ and $V(t)$ as follows

\begin{eqnarray}\phi(t) = \int^{t}_{0}\left[-\frac{2\dot{H}}{3H^{2}}\right]^{1/2}dt+\phi_{0}\label{n14}\end{eqnarray}
with $\phi_{0}$ as the value of $\phi$ at $t=0$,  and
\begin{eqnarray}V(t)=\frac{3H^{2}}{8\pi G}\left[1+\frac{2\dot{H}}{3H^{2}}\right]^{1/2}\label{n15}.\end{eqnarray}

Thus,  we can proceed to determine $V(\phi)$ for the cases (\ref{n10}) and (\ref{n11})  by     eliminating $t$ from (\ref{n14}) and (\ref{n15}). This provides us with different time dependence of the Hubble parameter in the two cases given by (\ref{n10}) and (\ref{n11}).

For the scale factor of type $a(t)=a_{0}t^{n}exp(\alpha t)$, where $a=a_{0}$ at the epoch $\tau$ such that $\tau^{n}exp(\alpha \tau)=1$, this gives
\begin{eqnarray}H(t)=nt^{-1}+\alpha;\quad \dot{H}=-nt^{-2}\label{x2}\end{eqnarray} and hence from (\ref{n14})
\begin{eqnarray}\phi(t)=\left(\frac{2n}{3}\right)^{1/2}\int^{t}_{0}\frac{1}{(n+\alpha t)}dt+\phi_{0}\label{n18}\end{eqnarray} with $\phi=\phi_{0}$ at $t=0$. Therefore,  an arbitrary epoch in such evolution of the scale factor is given by

\begin{eqnarray}t=-\frac{n}{\alpha}+\frac{n}{\alpha}exp\left[\alpha\sqrt{\frac{3}{2n}}(\phi(t)-\phi_{0})\right].\label{n19}\end{eqnarray}
Putting the forms of $H$ and $\dot{H}$ in equation (\ref{n15}), we get
\begin{eqnarray}V(t)=\frac{\sqrt{3}(n+\alpha t)}{8\pi G t^{2}}[3(n+\alpha t)^{2}+2n]^{1/2}.\label{n20}\end{eqnarray}
Thus, finally eliminating $t$ from (\ref{n18}) and (\ref{n20}), we get $V(\phi)$ as
\begin{eqnarray}V(\phi)=\frac{\sqrt{3} n exp\left(\alpha \sqrt{\frac{3}{2n}}(\phi(t)-\phi_{0})\right)\left[3n^{2}exp \left(2\alpha\sqrt{\frac{3}{2n}}(\phi(t)-\phi_{0})\right)+2n\right]^{1/2}}{8\pi G \left[-\frac{n}{\alpha}+\frac{n}{\alpha}exp\left(\alpha\sqrt{\frac{3}{2n}}(\phi(t)-\phi_{0})\right)\right]^{2}}.\label{n21}\end{eqnarray}
 As expected, it can be seen that $ V(\phi)\rightarrow\infty $ as $t\rightarrow 0 $ and $\phi(t)\rightarrow\phi_{0}$.

In the second case (\ref{n11}), we take the evolution of scale factor as $a(t)=At^{n}+B exp(\alpha t)$ in form of a linear combination of power law and exponential expansion. Then, the Hubble parameter becomes \begin{eqnarray}H=\frac{A n t^{n-1}+ B  \alpha exp (\alpha t)}{A t^{n}+ B exp(\alpha t)}\label{n22}\end{eqnarray}
and \begin{eqnarray}\dot{H}=\frac{A n(n-1)t^{n-2}+ B \alpha^{2}exp(\alpha t)}{At^{n}+ B exp(\alpha t)}-\left[\frac{A n t^{n-1}+ B \alpha exp(\alpha t)}{A t^{n}+ B exp(\alpha t)}\right]^{2}.\label{n23}\end{eqnarray}
Using the above  expressions in (\ref{n14}) and (\ref{n15}),  we obtain
\begin{eqnarray}\phi(t)=\int^{t}_{0}\left[-\frac{2}{3}\left(\frac{(A n(n-1)t^{n-2}+B\alpha^{2}exp(\alpha t))(A t^{n}+ B exp(\alpha t))}{(A n t^{n-1}+ B \alpha exp(\alpha t))^{2}}\right)\right]^{1/2}dt +\sqrt{\frac{2}{3}}t + \phi_{0}\label{n24}\end{eqnarray}
and \begin{eqnarray}V(t)=\frac{\sqrt{3}H^{2}}{8 \pi G}(A t^{n} + B exp(\alpha t))^{1/2}\left[1+\frac{2A n(n-1) t^{n-2}-2B\alpha^{2} exp(\alpha t)}{(A n t^{n-1}+B\alpha exp(\alpha t))^{2}}\right].\label{n25}\end{eqnarray}
In this way, by (\ref{n24}) and (\ref{n25}), we can evaluate $V(\phi)$ for the scale factor under consideration.

\section{Mutual interaction  between the  tachyonic scalar field components}

The interacting dark energy models have been recently proposed by several authors \cite{w1,x1,x2,x3,x4,x5}. Here we study the components of relativistic scalar field (tachyonic scalar field) as interacting mutually. Under these circumstances, the cosmological constant is no longer a true constant and the rate of its decline is ascertained by the complementary evolution of dark matter densities. It is obvious that during interaction the overall conservation of energy is kept intact.
The individual equations of energy conservation for dark matter and the cosmological constant are respectively given as

\begin{eqnarray}\dot{\rho_{m}}=-3H(1+w_{m})\rho_{m}+Q; \qquad \dot{\rho_{\lambda}}=-3H(1+w_{\lambda})\rho_{\lambda}-Q\label{x1}.\end{eqnarray}

With $w_{m}=0$  and  $w_{\lambda}=-1$,  these become

\begin{eqnarray}\dot{\rho_{m}}=-3H\rho_{m}+Q\label{n26}\end{eqnarray}  and  \begin{eqnarray}\dot{\rho_{\lambda}}=-Q.\label{n27}\end{eqnarray}

Several authors have proposed different forms of $Q$ \cite{y1,z1,z2,z3}. Due to the lack of information regarding the  exact nature of dark matter and dark energy (as the cosmological constant or else) we present the form of interaction term $Q$ heuristically. Here, we impose the following simple assumptions about $Q$ :

\textbf{(i)} $Q$ should be small and positive.  If it had large and negative value then dark energy would have dominated the expansion practically from the outset and galaxies could not have formed at the desired epochs.

\textbf{(ii)} It is linear combination of $\dot{\rho_{m}}$ and $\dot{\rho_{\lambda}}$  that is, $Q=\gamma\dot{\rho_{m}}+\beta\dot{\rho_{\lambda}}$, where $\gamma$ and $\beta$ are the proportionality constant whose values may be determined by the observations as discussed in section (5). Any time dependence, that most of authors include in the interaction through Hubble parameter, has been automatically included in our assumption via the evolution of $\rho_{m}$ and $\rho_{\lambda}$. Thus,  there is no use of $H$  as such in $Q$.
Therefore, there arise the following four cases:
\begin{enumerate}
\item[(1)]  $\gamma\neq0$, $\beta=0$ and \textbf{$Q=\gamma \dot{\rho_{m}}$}.

\item[(2)]  $\beta\neq0$, $\gamma=0$ and \textbf{ $Q=\beta\dot{\rho_{\lambda}}$}.

\item[(3)]  $\gamma\neq0$, $\beta\neq0$, $\gamma=\beta=\sigma$, \textbf{$Q=\sigma\dot{\rho}$}.

\item[(4)]  $\gamma\neq\beta\neq0$, \textbf{$Q=\gamma \dot{\rho_{m}}+\beta\dot{\rho_{\lambda}}$}.
\end{enumerate}

Now,  we determine the evolution of $\rho_{m}$ and $\rho_{\lambda}$ in each case.  First we discuss \textbf{$Q=\gamma \dot{\rho_{m}}$}. From (\ref{n26}) we have $(1-\gamma)\dot{\rho_{m}}/\rho_{m}=-3\dot{a}/a$.  Integrating this,  we get

\begin{eqnarray}\rho_{m} =\rho^{0}_{m}\left(\frac{a}{a_{0}}\right)^{-3/(1-\gamma)}\label{n28}\end{eqnarray} where $\rho^{0}_{m}$ and $a_{0}$ are the values of dark matter energy density and the scale factor respectively, both at the present epoch.
The evolution of $\lambda$ energy density can be determined from (\ref{n27}) as
\begin{eqnarray}\dot{\rho_{\lambda}}=-Q=-\gamma \dot{\rho_{m}}\label{k1}.\end{eqnarray}
Using the time derivative of (\ref{n28}) and integrating (\ref{k1}),  we get
\begin{eqnarray}\rho_{\lambda}=\rho^{0}_{\lambda} - \gamma \rho^{0}_{m}\left[\left(\frac{a}{a_{0}}\right)^{-3/1-\gamma}-1\right].\label{n30}\end{eqnarray}
It can be seen that in the absence of interaction ($\gamma=0$) we have $\rho_{m}=\rho^{0}_{m}(a/a_{0})^{-3}$ and  $\rho_{\lambda}=\rho^{0}_{\lambda}=$ constant, as expected in the standard approach with a truly constant cosmological constant.

Considering the interaction term \textbf{$Q=\beta\dot{\rho_{\lambda}}$} from the energy conservation law for the cosmological constant (\ref{n27}), we obtain $(1+\beta)\dot{\rho_{\lambda}}=0$.  Here, three possibilities  arise:

\begin{enumerate}
\item[(a)] $1+\beta \neq 0$, $\rho_{\lambda}=$ constant

\item[(b)] $1+\beta=0$,  $\rho_{\lambda}=$ constant and

\item[(c)] $1+\beta=0$,  $\dot{\rho_{\lambda}}\neq 0$ and then  $\beta=-1$ and a time varying energy density of $\lambda$ component will yield the desired interaction.
\end{enumerate}
 However, $Q=\beta\dot{\rho_{\lambda}}$ is a redundant constraint for interaction.  The reason lies in the fact that out of the above three conditions, the first two do not give any interaction at all, whereas the third one implies $\beta=-1$ and so $Q=-\dot{\rho_{\lambda}}$ which has been already satisfied  in case of interaction $Q=\gamma\dot{\rho_{m}}$.

Next, we consider the interaction term of the form \textbf{$Q=\sigma\dot{\rho}=\sigma(\dot{\rho_{m}}+\dot{\rho_{\lambda}})$}.  Then (\ref{n26}) and (\ref{n27}) give
$\dot{\rho_{\lambda}}=-\sigma(\dot{\rho_{m}}+\dot{\rho_{\lambda}})$. On its integration,  we obtain
\begin{eqnarray}\rho_{\lambda}=\rho^{0}_{\lambda}+\frac{\sigma}{1+\sigma}\rho^{0}_{m}-\frac{\sigma}{1+\sigma}\rho_{m}\label{n31} \end{eqnarray}
while for matter, we have $\dot{\rho_{m}}=\sigma(\dot{\rho_{m}}+\dot{\rho_{\lambda}})$
\begin{eqnarray}\frac{\dot{\rho_{m}}}{\rho_{m}}=-3(1+\sigma)\frac{\dot{a}}{a}\label{m3}\end{eqnarray}
which can be integrated to lead to the solution
\begin{eqnarray}\rho_{m}=\rho^{0}_{m}\left(\frac{a}{a_{0}}\right)^{-3(1+\sigma)}\label{n32}.\end{eqnarray}

Here, the striking consequence of (\ref{n31}) is that $\rho_{\lambda}$ goes down linearly with $\rho_{m}$, the later depending on the scale factor according to (\ref{n32}).

Considering the interaction term \textbf{$Q=\gamma\dot{\rho_{m}}+\beta\dot{\rho_{\lambda}}$}, and from the energy conservation law (\ref{n27}),  we have
\begin{eqnarray}\dot{\rho_{\lambda}}=-\frac{\gamma}{1+\beta}\dot{\rho_{m}}\label{m4}\end{eqnarray} whose solution is given by
\begin{eqnarray}\rho_{\lambda}=\rho^{0}_{\lambda} +\frac{\gamma}{1+\beta}\rho^{0}_{m} -\frac{\gamma}{1+\beta}\rho_{m}\label{n33}\end{eqnarray} and for matter, we have
\begin{eqnarray}\frac{\dot{\rho_{m}}}{\rho_{m}}=-\frac{3(1+\beta)}{(1+\beta-\gamma \beta)}\frac{\dot{a}}{a}\label{n34}.\end{eqnarray}
Integrating (\ref{n34}),  the functional form of $\rho_{m}$ is obtained as
\begin{eqnarray}\rho_{m}=\rho^{0}_{m}\left(\frac{a}{a_{0}}\right)^{-\frac{3(1+\beta)}{(1+\beta-\gamma \beta)}}\label{n35}.\end{eqnarray}

Thus, for various possibilities of interaction terms, we have different forms of variation of $\rho_{m}$ and $\rho_{\lambda}$ with the evolving scale factor in the expanding universe. The exact values of the parameters $\gamma$, $\beta$ or $\sigma$,   as may be determined by the observations, are discussed in section (5).

\section{\textbf{$O_{m}(x)$ diagnostic for the interacting cosmological constant }}
The $O_{m}(x)$  diagnostic proposed earlier \cite{a2}, and significantly including the effect of interaction, is given by

\begin{eqnarray}O_{m}(x)=\frac{E^{2}(x)-1}{x^{3}-1}\label{n36}\end{eqnarray}

Here $x =1+z=a_{0}/a$, and $E^{2}(x) =H^{2}(x)/H^{2}_{0}$. From the Friedmann equations (\ref{n12}) for spatially flat geometry, we have
\begin{eqnarray}H^{2}=\frac{8\pi G}{3}(\rho_{m}+\rho_{\lambda})\label{n37}\end{eqnarray}
For interaction term $Q=\gamma\dot{\rho_{m}}$,
\begin{eqnarray}E^{2}(x)=\frac{H^{2}(x)}{H^{2}_{0}}=\Omega^{0}_{m}x^{3/1-\gamma}+\Omega^{0}_{\lambda}-\gamma\Omega^{0}_{m}\left(x^{3/1-\gamma}-1\right)\label{n38}.\end{eqnarray}
Substituting the value of $E^{2}(x)$ in $O_{m}(x)$ from (\ref{n38}) in (\ref{n36}) and using $\Omega^{0}_{\lambda}=1-\Omega^{0}_{m}$ for the spatially flat universe, we have
\begin{eqnarray}O_{m}(x)=\Omega^{0}_{m}(1-\gamma)\left(\frac{x^{3/1-\gamma}-1}{x^{3}-1}\right)\label{n39}.\end{eqnarray}
In the absence of interaction $\gamma=0$, $O_{m}(x)=\Omega^{0}_{m}$ as expected.

For the  interaction term $Q=\sigma\dot{\rho}$,  $O_{m}(x)$ becomes

\begin{eqnarray}O_{m}(x)=\frac{\Omega^{0}_{m}}{1+\sigma}\left[\frac{x^{3(1+\sigma)}-1}{x^{3}-1}\right]\label{n40}\end{eqnarray}

while for the interaction term $Q=\gamma\dot{\rho_{m}}+\beta\dot{\rho_{\lambda}}$,  the above diagnostic adopts the form

\begin{eqnarray}O_{m}(x)=\frac{\Omega^{0}_{m}}{1+\beta}\left[\frac{(1+\beta-\gamma)x^{3(1-\beta)/1+\beta-\gamma\beta}+\gamma-1-\beta}{x^{3}-1}\right]\label{n41}.\end{eqnarray}

Here,  again it is clear that in the absence of interaction,  (\ref{n39}), (\ref{n40}) and (\ref{n41}) give
$O_{m}(x)=\Omega^{0}_{m}$. Thus,  we see that the interaction coefficients appear in $O_{m}(x)$ and so the corresponding observations will effectively constrain the interaction between the components of the tachyonic field.

\section{\textbf{Determination of the proportionality constants in different  forms of $Q$ interaction}}

In this section, we  discuss a possible approach to calculate the proportionality constants for different forms of interaction $Q$ through the observations of Hubble parameter at different redshifts $z$ along the evolution of the universe.

The difference of the squares of the normalized Hubble parameter $E^{2}(x)$ at two different redshifts $x_{i}$  and $x_{j}$ for the interaction term $Q=\gamma\dot{\rho_{m}}$ is given as
\begin{eqnarray} E^{2}(x_{i})-E^{2}(x_{j})=\triangle E^{2}(x_{i},x_{j})=(1-\gamma)\Omega^{0}_{m}\left[x^{3/1-\gamma}_{i}-x^{3/1-\gamma}_{j}\right]\label{n42}.\end{eqnarray}

Similarly, $\triangle E^{2}(x_{i},x_{j})$ can be found  for  $Q=\sigma\dot{\rho}$ and $Q=\gamma\dot{\rho_{m}}+\beta\dot{\rho_{\lambda}}$ as

\begin{eqnarray}\triangle E^{2}(x_{i},x_{j})=\frac{\Omega^{0}_{m}}{1-\sigma}\left[x^{3(1+\sigma)}_{i}-x^{3(1+\sigma)}_{j}\right].\label{n43} \end{eqnarray} and  \begin{eqnarray}\triangle E^{2}(x_{i},x_{j})=\frac{(1+\beta-\gamma)\Omega^{0}_{m}}{1+\beta}\left[x^{3(1-\beta)/1+\beta-\gamma\beta}_{i}-x^{3(1-\beta)/1+\beta-\gamma\beta}_{j}\right]\label{n44}\end{eqnarray} respectively.

Measuring the values of $\triangle E^{2}(x_{i},x_{j})$ from the redshift observations and $\Omega^{0}_{m}$ as  the concordance value  from  CMBR, BAO etc. data analysis,  we calculate $\gamma$, $\sigma$ and  $\beta$ for the  $(x_{i}, x_{j})$ pair of redshifts corresponding to the selected interaction forms. We emphasize  that the choice of such pair is desirable over a large range of redshift observations,  since we must determine the interaction strength over a long period of cosmic evolution. This is required so as to  constrain the structure formation both at high- and low-$z$ epochs. The precise measurements of $H_{0}$ will be helpful to break the degeneracy among the cosmological parameters \cite{f2}. Here,  we take a set of values of $H(z)$  \cite{h2,i2} as:  at $z=0.1$, $0.4$, $1.3$, and $1.75$,  the values of $H(z)=$ $69\pm12$, $95\pm17$, $168\pm17$ and $202\pm40$ km s$^{-1}$  Mpc$^{-1}$ respectively. Further calculations  may be done by taking the present values  $H_{0}=73.8\pm2.4 \approx 73.8 $ km s$^{-1}$  Mpc$^{-1}$ \cite{i3}  and the density parameter of matter component  $\Omega^{0}_{m}=0.272$  \cite{l2}.

Thus,  choosing the four epochs as $x_{1}=1.1$,\quad $x_{2}=1.4$,\quad $x_{3}=2.3$ and $x_{4}=2.75$ for these redshifts, the squared normalized Hubble parameter may be calculated in a straightforward  manner as given below:

$E^{2}(x_{1}) = 0.874$,\quad $E^{2}(x_{2}) = 1.657$,\quad $E^{2}(x_{3})=5.182$, and $E^{2}(x_{4})=7.492$.

and thus $\Delta E^{2}(x_{1},x_{2})=-0.783$,\quad $\Delta E^{2}(x_{1},x_{3})=-4.308$,

\quad $\Delta E^{2}(x_{1},x_{4})=-6.618$,\quad $\Delta E^{2}(x_{2},x_{3})=-3.525$,

\quad $\Delta E^{2}(x_{2},x_{4})=-5.835$ and $\Delta E^{2}(x_{3},x_{4})=-2.310$

Using these data we can proceed to  determine six values of the proportionality constant $\gamma$ of the interaction form $Q=\gamma\dot{\rho_{m}}$ from (\ref{n42})

$(1-\gamma)[(1.1)^{3/1-\gamma}-(1.4)^{3/1-\gamma}]   =-2.879$,

$(1-\gamma)[(1.1)^{3/1-\gamma}-(2.3)^{3/1-\gamma}]   =-15.838$,

$(1-\gamma)[(1.1)^{3/1-\gamma}-(2.75)^{3/1-\gamma}]  =-24.330$,

$(1-\gamma)[(1.4)^{3/1-\gamma}-(2.3)^{3/1-\gamma}]   =-12.960$,

$(1-\gamma)[(1.4)^{3/1-\gamma}-(2.75)^{3/1-\gamma}]  =-21.452$, and

$(1-\gamma)[(2.3)^{3/1-\gamma}-(2.75)^{3/1-\gamma}]  =-8.493$.

In the similar way,  $\sigma$  and $\beta$   can be calculated from (\ref{n43}) and (\ref{n44}) respectively as follows :

$(\frac{1}{1-\sigma})[(1.1)^{3(1+\sigma)}-(1.4)^{3(1+\sigma)}]=-2.879$,

$(\frac{1}{1-\sigma})[(1.1)^{3(1+\sigma)}-(2.3)^{3(1+\sigma)}]=-15.838$,

$(\frac{1}{1-\sigma})[(1.1)^{3(1+\sigma)}-(2.75)^{3(1+\sigma)}]=-24.330$,

$(\frac{1}{1-\sigma})[(1.4)^{3(1+\sigma)}-(2.3)^{3(1+\sigma)}]=-12.960$,

$(\frac{1}{1-\sigma})[(1.4)^{3(1+\sigma)}-(2.75)^{3(1+\sigma)}]=-21.452$,

$(\frac{1}{1-\sigma})[(2.3)^{3(1+\sigma)}-(2.75)^{3(1+\sigma)}]=-8.493$,

and

$(\frac{1+\beta-\gamma}{1+\beta})[(1.1)^{3(1-\beta)/1+\beta-\gamma}-(1.4)^{3(1-\beta)/1+\beta-\gamma}]=-2.879$,

$(\frac{1+\beta-\gamma}{1+\beta})[(1.1)^{3(1-\beta)/1+\beta-\gamma}-(2.3)^{3(1-\beta)/1+\beta-\gamma}]=-15.838$,

$(\frac{1+\beta-\gamma}{1+\beta})[(1.1)^{3(1-\beta)/1+\beta-\gamma}-(2.75)^{3(1-\beta)/1+\beta-\gamma}]=-24.330$,

$(\frac{1+\beta-\gamma}{1+\beta})[(1.4)^{3(1-\beta)/1+\beta-\gamma}-(2.3)^{3(1-\beta)/1+\beta-\gamma}]=-12.960$,

$(\frac{1+\beta-\gamma}{1+\beta})[(1.4)^{3(1-\beta)/1+\beta-\gamma}-(2.75)^{3(1-\beta)/1+\beta-\gamma}]=-21.452$,

$(\frac{1+\beta-\gamma}{1+\beta})[(2.3)^{3(1-\beta)/1+\beta-\gamma}-(2.75)^{3(1-\beta)/1+\beta-\gamma}]=-8.493$.

 We will further calculate in another paper the values of  different proportionality constants over a wider  range of observations compatible with the structure formation at the corresponding epochs.

\section{Conclusion}

Having the motivation for the  relativistic (tachyonic) scalar field, in contrast to quintessence, and  in the  absence of our knowledge hitherto of any definite form of the evolution of the scale factor in the presently accelerating universe, in section 2  we took the choice of the linear combination of the power law and exponential expansion.  This was further used to find the potential  $V(\phi)$ in each case. Since it is hard to believe that the cosmological constant does drive the present acceleration yet is left with  no interaction with the background, we studied in section 3  various possible  forms of interaction between the two components of the tachyonic field -- matter and the cosmological constant. We determined the evolution of the components of matter and the cosmological constant.  In section 4, we showed that the previously given $O_{m}(x)$  diagnostic  effectively reflects any such interaction.  We  further showed  that the correct form of the  interaction can be determined by the observations at high and low  redshifts, over a wide range to constrain the structure formation at those epochs.  This sets pace for the future study in matching such observations with the theoretical predictions about the interaction to narrow down its form.

\section{Acknowledgment}

The authors are thankful to the University Grants Commission, New Delhi for its support for the present work  through the Major Research Project vide  F. No.37-431/2009 (SR).

\end{document}